\documentclass[runningheads]{llncs}

\usepackage[T1]{fontenc}
\def\doi#1{\href{https://doi.org/\detokenize{#1}}{\url{https://doi.org/\detokenize{#1}}}}
\usepackage{booktabs}
\usepackage{multirow}
\usepackage{graphicx, color, soul}

\usepackage[misc,geometry]{ifsym}
\usepackage{hyperref}
\usepackage{listings}
\begin{document}
%
%\title{Landmark-based Automatic Ultrasound Fetal Biometry}
\title{BiometryNet: Landmark-based Fetal Biometry Estimation from Standard Ultrasound Planes}
\titlerunning{BiometryNet: Landmark-based Fetal Biometry Estimation}
% \author{*****************}
% \institute{**********}

\author{Netanell Avisdris \textsuperscript{\Letter} \inst{1,2} \orcidID{0000-0002-0719-1327}  \and %index{Avisdris, Netanell}
Leo Joskowicz\inst{1} \and %index{Joskowicz, Leo}
Brian Dromey\inst{4,6} \and %index{Dromey, Brian}
Anna L. David\inst{6} \and %index{David, Anna L.}
Donald M. Peebles\inst{6} \and %index{Peebles, Donald M.}
Danail Stoyanov\inst{4,5} \and %index{Stoyanov, Danail}
Dafna Ben Bashat\inst{2,3} \and %index{Ben Bashat, Dafna}
Sophia Bano\inst{4,5} } %index{Bano, Sophia}

\authorrunning{N. Avisdris et al.}
\institute{School of Computer Science and Engineering, The Hebrew U. of Jerusalem, Israel \email{\{netana03,josko\}@cs.huji.ac.il} \and
Sagol Brain Institute, Tel Aviv Sourasky Medical Center, Israel \and Sagol School of Neuroscience and Sackler Faculty of Medicine, Tel Aviv U., Israel \and
Wellcome/EPSRC Centre for Interventional and Surgical Sciences (WEISS),
University College London, UK \and
Department of Computer Science, University College London, UK \and
Elizabeth Garrett Anderson Institute for Women’s Health, U. College London, UK
}

\maketitle              % typeset the header of the contribution
\begin{abstract}
Fetal growth assessment from ultrasound is based on a few biometric measurements that are performed manually and assessed relative to the expected gestational age. Reliable biometry estimation depends on the precise detection of landmarks in standard ultrasound planes. Manual annotation can be time-consuming and operator dependent task, and may results in high measurements variability. Existing methods for automatic fetal biometry rely on initial automatic fetal structure segmentation followed by  geometric landmark detection. However, segmentation annotations are time-consuming and may be inaccurate, and landmark detection requires developing measurement-specific geometric methods. 
This paper describes BiometryNet, an end-to-end landmark regression framework for fetal biometry estimation that overcomes these limitations. It includes a novel Dynamic Orientation Determination (DOD) method for enforcing measurement-specific orientation consistency during network training. DOD reduces variabilities in network training, increases landmark localization accuracy, thus yields accurate and robust biometric measurements.
To validate our method, we assembled a dataset of 3,398 ultrasound images from 1,829 subjects acquired in three clinical sites with seven different ultrasound devices. Comparison and cross-validation of three different biometric measurements on two independent datasets
shows that BiometryNet is robust and yields accurate measurements whose errors are lower than the clinically permissible errors, outperforming other existing automated biometry estimation methods. Code is available at \href{https://github.com/netanellavisdris/fetalbiometry}{https://github.com/netanellavisdris/fetalbiometry}.

% The abstract should briefly summarize the contents of the paper in

\end{abstract}
\section{Introduction}
Ultrasound (US) based estimation of fetal biometry is widely used to monitor fetal growth assessed relative to the expected gestational age and to diagnose prenatal abnormalities. Reliable estimation of fetal biometry depends on the localization of fetal landmarks on standard planes (SPs) by the sonographer. Common SPs include the trans-ventricular plane to measure the fetal head circumference, the trans-abdominal plane to measure the fetal abdominal circumference, and femoral plane to measure the femur length. 
Obtaining manual measurements can be time-consuming and operator-dependent, especially for trainees~\cite{dromey2020dimensionless}, and results in high inter-and intra-operator variability~\cite{joskowicz2019observervar}. In fetal biometry, inter-operator measurements variability range between ±4.9$\%$ and ±11.1$\%$ and an intra-operator variability between ±3$\%$ and ±6.6$\%$ \cite{sarris2012intra}. This variability contributes to biometry uncertainty and hampers fetal growth evaluation in the clinic. 
Automating fetal biometry can reduce measurement variability and improve fetal monitoring and clinical decision making. 

Numerous methods have been developed for automatic computation of different measurements in US-based fetal biometry \cite{torres2022headus_review}, e.g., biparietal diameter \cite{al2019bpdauto,zhang2017automatic_fhc}, head circumference \cite{van2018automated}, and femur length \cite{khan2017bpdfl}. Recently, methods for comprehensive biometry required for fetal weight estimation have been proposed~\cite{bano2021autofb,prieto2021automated}. Similar methods have been developed for other imaging modalities, e.g., fetal MRI \cite{avisdris2021automatic} and 3D US \cite{ryou2019fetal3dus}. These methods rely on obtaining segmentation masks to train fetal anatomy segmentation models, which is  time-consuming, and on developing geometric methods for each measurement. \looseness=-1

An alternative approach is the detection of landmarks directly on the image without relying on segmentation masks. The advantages of this approach are that it follows the clinical workflow -- the sonographer locates on the two landmarks required for biometric measurement directly on the image, and that obtaining ground-truth data is fast and straightforward. Landmark-based approaches have been proposed for various medical imaging modalities \cite{zhang2017t2dl}, e.g. CT and MRI, to quantify lesions progression \cite{tang2021lesion}, for fetal MRI biometry estimation \cite{avisdris2021fmlnet}, and for fetal landmark detection in US using reinforcement learning agents \cite{alansary2019evaluating}. However, this approach has not been used for automating fetal biometry in US SPs. \looseness=-1

\begin{figure}[tp]
\includegraphics[width=\textwidth]{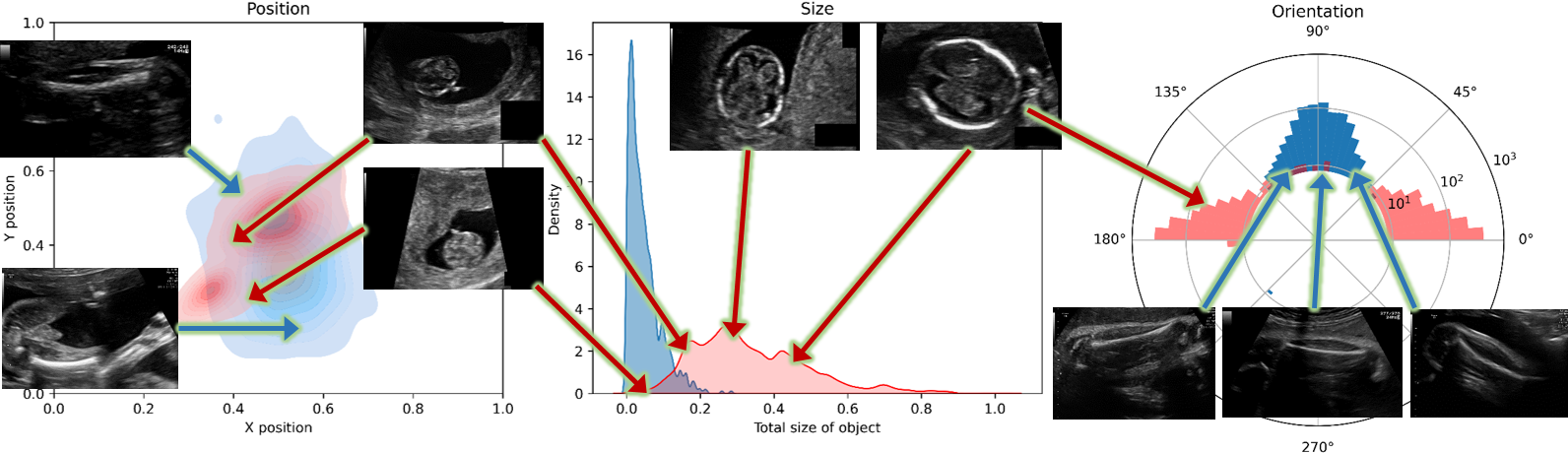}
\caption{Fetal biometry variability for the head (red) and femur (blue) SPs. The graphs show (left to right) the distributions of the position (structure center point), size (structure area with respect to image area) and orientation (structure angle with respect to horizontal plane). Sample US image inserts illustrate the distribution.\looseness = -1} \label{fig_dboreint}
\end{figure}

% Challanges paragraph: Orientation, scale, annotation of points vs segmentation - BD "I think this is great paragraph, super important to extablish why this is so difficult!",
Automating fetal US biometry presents numerous challenges. First, imaging quality varies across different US devices, heterogeneous in overall appearance, particularly regarding gain and zoom. Second, some fetal anatomical structures may become difficult to image with increasing gestational age. Third, anatomical structure position, size, orientation, and appearance present significant variability (Fig. \ref{fig_dboreint}). Since the fetus can lie in any of a wide variety of positions, anatomical landmarks can be in orientations that are off the horizontal.
%inter and intra-observer variabilities,
%some of these orientations may compromise the SP which can be achieved clinically.
 %; contributing further to image heterogeneity. 
Finally, biometric parameters have different geometric characteristics and performed on different SPs, thus each requires its own individual method. 
% To be useful, an automatic fetal biometry system must address these challenges to achieve robustness.
% Any automated biometry system should consider incorporating these challenges for achieving generalizability.
\looseness=-1
%Most fundamental challenges include (a) large variability in the scanning planes, fetal brain size and orientation (Fig. \ref{fig_dboreint}), (b) different geometric characteristics of each biometry, (c) heterogeneity in US images in terms of overall appearance, gain and zoom selections and (d) added variability introduced by increasing gestation. These must be addressed for achieving generalizability in automating fetal biometry. 
%Any automated biometry system should consider incorporating these challenges for achieving generalizability. 
%Fundamentally, these clinical limitations present in one of three ways; which must be considered for any automated biometry system, such as we propose: (a) there is a large variability in the scanning planes, fetal brain size and orientation (Fig. \ref{fig_dboreint}). (b) each biometry have different geometric characteristics, (c) US images of the fetus are heterogeneous in overall appearance, particularly gain and zoom selections, and this can be further influenced by gestation. 

We propose BiometryNet, an end-to-end framework that automates US fetal biometry for multiple fetal structures using direct landmark detection that overcomes these challenges. Unlike other methods ~\cite{bano2021autofb,al2019bpdauto,ryou2019fetal3dus,zhang2017automatic_fhc,torres2022headus_review}, BiometryNet only requires measurement landmark annotations for training. 
% without the need of any geometric fitting as post-processing. 
The main contributions of our work are:
%(1) We show for the first time, that using landmark annotations alone, without segmentation masks, which are simpler to annotate, US fetal biometry can be automated with high accuracy; (2) We propose a novel Dynamic Orientation Determination (DOD) mechanism, which determine measurement-wise orientation and provide consistent landmark class for various measurements. We show that DOD is robust against variabilities in the SPs orientations, leading to improved biometry estimates; (3) We utilize and collect a large fetal biometry dataset, acquired from various clinical centers, machines and operators; and (4) We demonstrate and analyze BiometryNet generalization capabilities by performing cross-validation on two independent dataset;
\begin{itemize}
\item We show for the first time that US fetal biometry can be automated with high accuracy using only landmark annotations that can be acquired easily. We demonstrate our approach on a large fetal biometry dataset of 3,398 US images from 1,820 subjects acquired by various operators and US devices at multiple clinical centers. 
%\[0.1ex]
\vspace*{1.0ex}
\item We propose a novel Dynamic Orientation Determination (DOD) method to determine measurement-wise orientation and provide consistent landmark class for various measurements. We show that DOD is robust for variabilities in the SPs orientations, leading to improved biometry estimates and generalization on unseen datasets.
 
\end{itemize}

\section{Method}
\begin{figure}[tp]
\includegraphics[width=\textwidth]{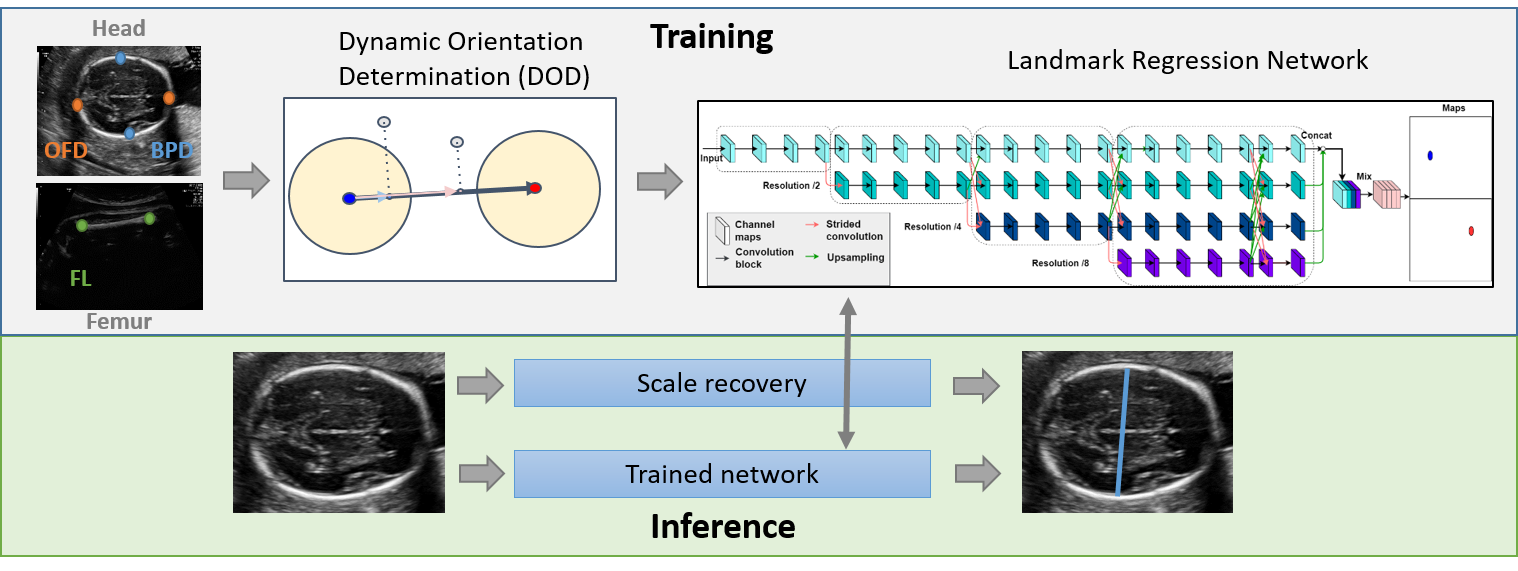}
\caption{BiometryNet framework. Top: during training, annotated head and femur planes are fed into the DOD module and to the landmark regression network to predict two landmarks per biometric measurement. Bottom: during inference, the trained model predicts the landmarks followed by scale recovery for biometric measurements estimation.} \label{fig_pipeline}
\end{figure}

We propose BiometryNet, a framework for the estimation of fetal US biometry using a landmark regression convolution neural network (Fig. \ref{fig_pipeline}). BiometryNet locates two landmark points on an US image for three biometric parameters: biparietal diameter (BPD) and occipito‐frontal diameter (OFD) in the trans-ventricular (head) plane, and femur length (FL) in the femur plane. It includes Dynamic Orientation Determination (DOD) used during training for consistent landmark class. 
%for various measurements, resulting in improved landmark localization. 
At inference time, the trained BiometryNet predicts two biometry landmarks, followed by scale recovery in order to estimate image resolution needed to compute the actual biometric measure.

\subsection{Landmark Regression Network}
The landmark regression network is a modified HRNet \cite{WangSCJDZLMTWLX19}, trained to predict a heat map for each landmark defined by a Gaussian function centered at the landmark coordinates whose co-variance describes the landmark location uncertainty. We used HRNet since it has achieved state-of-the-art performance for computer vision tasks, e.g. object detection, semantic segmentation, face landmark detection and human pose estimation. HRNet is a Convolutional Neural Network that combines the representations of multi-scale high-to-low resolution parallel streams into a single stream. The representations are then input to a two-layer convolution classifier. The first layer combines the feature maps of all four resolutions; the second layer computes a Gaussian heat map for each of the two landmarks. One network is trained for each biometric measurement (two landmarks points) with the Mean Squared Error loss between the Gaussian maps created from the ground-truth landmarks and the predicted heat maps. At inference time, the two landmarks' locations are defined by the coordinates of the pixel with the maximal value on each heat map.

To compensate for the high variability in acquisition orientation and object scales in fetal US images (Fig. \ref{fig_dboreint}), two training time augmentations are used: 1) rotations around the image center at randomly sampled angles in the range $[-180, 180]^o$ followed by cropping to preserve a fixed image size; 2) image scaling at randomly sampled scales in the range $[-5, +5]\%$. 

\subsection{Dynamic Orientation Determination (DOD)}
Since fetal structure may appear in various orientations (Fig.  \ref{fig_dboreint}) and CNN are not rotation-invariant, orientation variability must be handled properly. A common approach to handle such variability is augmentation. However, rotation augmentation may cause landmark class labeling (e.g left/right landmarks) to be inconsistent with image coordinates, i.e. the left and right points may be erroneously switched, which will hamper network training (Fig. 1 Supp.). This inconsistency can be corrected by landmark class reassignment (LCR) \cite{avisdris2021fmlnet} to preserve horizontal (left/right) landmark class consistency after augmentation. 
%LCR has been shown to improve biometry estimation accuracy.
% Landmark class reassignment~\cite{avisdris2021fmlnet} determines the specific class of horizontal (left-right) measurement landmark pairs within network training. This mechanism has been shown to improve biometry measurement using landmark regression network~\cite{avisdris2021fmlnet}, as it preserves landmark class consistency after all augmentations. 
%
However, different biometric measurements may have different spatial orientation, e.g., OFD is mostly vertical and BPD is mostly horizontal.

\begin{figure}[tp]
\includegraphics[width=\textwidth]{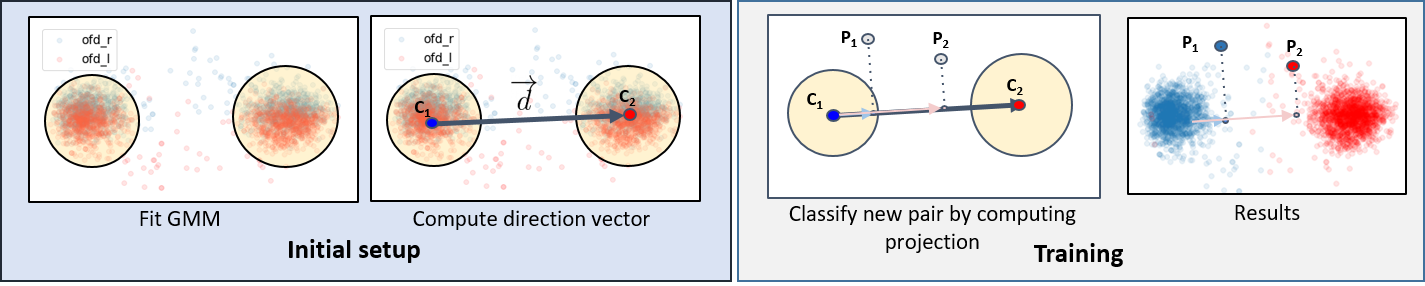}
\caption{Illustration of Dynamic Orientation Determination (DOD). Left: \it{Initial setup} stage, measurement direction vector computation between GMM centroids fitted to normalized landmark points; Right: \it{Training} stage, re-assignment of each landmark pair class by projecting the landmarks on the direction vector and ordering them.} \label{fig_orient}
\end{figure}

To overcome this issue, we introduced Dynamic Orientation Determination (DOD), a method that determine measurement-wise orientations and perform class reassignment accordingly, instead of computing only the horizontal measurement landmark pairs as in~\cite{avisdris2021fmlnet}.
DOD consists of two stages (Fig. \ref{fig_orient}). 
In \textit{initial setup} stage, biometry orientation is determined. First, Gaussian Mixture Model of two Gaussians was fitted onto the ground-truth landmarks of training dataset of each biometric measurements with Expectation-Maximization algorithm \cite{dempster1977EM}. Next, biometry orientation is computed as directional vector between the two Gaussian centroids, $\overrightarrow{d} = \overrightarrow{C_1C_2}$. 
In the network \textit{training} stage, the learned orientation is used to enforce consistency. After all augmentations are performed, each resulting biometric measurement landmark pair ($P = p_1, p_2$) is projected on the directional vector $r_i = (p_i \cdot \overrightarrow{d})/{|\overrightarrow{d}|}$ and then ordered according to its projection $sort(|r_i|)$ to obtain their reassigned class. 

\subsection{Scale Recovery}
To obtain actual measurements that can be compared to those obtained in the clinic, scale conversion from pixel to millimeter units is required. While this information is usually available during examination or is embedded in the raw image data, some retrospectively collected images may lack it. Therefore, we perform scale recovery using the approach presented in~\cite{bano2021autofb}. Briefly, it recovers true scale by detecting ruler markers using template matching.

\section{Experimental Setup}

{\bf Datasets and Annotations:}
we use two public datasets, Fetal Planes (FP)~\cite{burgos2020zdataset} and HC18~\cite{van2018automated}. 
FP \cite{burgos2020zdataset} was originally designed for US SPs classification challenge. The US images were acquired on six US devices: three GE Voluson E6, one Voluson S8, one Voluson S10 and one Aloka at two clinical sites in Barcelona, Spain. Since not all images in FP qualified as SPs for fetal biometry~\cite{salomon2008score}, we selected 1,638 (909 subjects) fetal head and 761 (630 subjects) fetal femur SPs. Not all subjects are included in both planes datasets, resulting in a total of 2,399 images from 1,014 unique subjects. An obstetrician then manually annotated the landmarks on each image with the VIA annotation tool \cite{dutta2019via}. On average, each plane annotation for landmarks took 20 seconds, which is far less than 70 seconds required for manual structure delineation.
HC18 \cite{van2018automated} was designed for fetal head circumference (HC) challenge. The US images were acquired with two US devices: GE Voulson E8 and 730, in one clinical site in the Netherlands. All HC18 training set, 999 (806 subjects) fetal head SPs, were annotated with a HC measurement. We computed the BPD and OFD biometric measurements from the major and minor axes of an ellipse by least-square fitting~\cite{fitzgibbon1996ellipsefit} onto the ground-truth mask. \looseness=-1 
\\[2ex]
{\bf Evaluation Metrics:}
we use the mean and median of $L_1$ difference, bias and agreement. For two sets of $n$ biometric measurements, $M_1=\{m_i^1\}$, $M_2=\{m_i^2\}$, let $m_i^1$ and $m_i^2$ ($ 1 \leq i \leq n$) be two measurement values, the ground-truth and the computed one, respectively.
The difference between each pair of measurements is defined as $d_i = m_i^1 - m_i^2$. The mean and median differences for $M_1, M_2$ are defined as $\overline{L_{1}}(M_1, M_2) = 1/n\sum_{i=1}^{n}{|d_i|}$ and $\widetilde{L_{1}}(M_1, M_2) = median_i({|d_i|})$, respectively. We use the Bland-Altman method \cite{altman1983measurement} to estimate the bias and agreement between two biometry measurement sets. Agreement is defined by the 95$\%$ confidence interval $CI_{95}(M_1, M_2) = 1.96 \times \sqrt{1/n\sum_{i=1}^{n}{(\overline{L_{1}}(M_1, M_2) - d_i)^2}}$. The measurements bias is defined as $Bias(M_1,M_2) = 1/n\sum_{i=1}^{n}{d_i}$.
\\[2ex]
{\bf Study Setup:}
the two datasets FP and HC18 were split into training and test sets. For the FP head and femur SPs, we selected 757 (449 subjects) and 437 (368 subjects) images for training and 881 (460 subjects) and 324 (262 subjects) images for testing, respectively. 
For a fair comparison of head planes in FP and HC18, we selected from HC18, similar number of 737 images (600 subjects) for training. The remaining 262 images (206 subjects) were used for the test set.
BiometryNet was implemented in PyTorch and trained for 200 epochs in about 2 hours on a single NVIDIA 1080Ti GPU with a batch size of 16 using the ADAM optimizer~\cite{KingmaB14}, an initial learning rate of $10^{-4}$ and a drop factor of 0.2 in epochs 10, 40, 90, and 150. For validation, a comparison to four networks was performed: (a) Vanilla HRNet \cite{WangSCJDZLMTWLX19}; (b) HRNet with fixed horizontal orientation determination; (c) HRNet with fixed vertical orientation determination for biometric measurements with horizontal orientation (OFD, FL) and vertical orientation (BPD); and, (d) FMLNet \cite{avisdris2021fmlnet}, which is an HRNet with fixed horizontal orientation determination, test-time augmentation and a method for prediction reliability estimation. 
\looseness=-1

\section{Results}
\begin{figure}[tp]
\includegraphics[width=\textwidth]{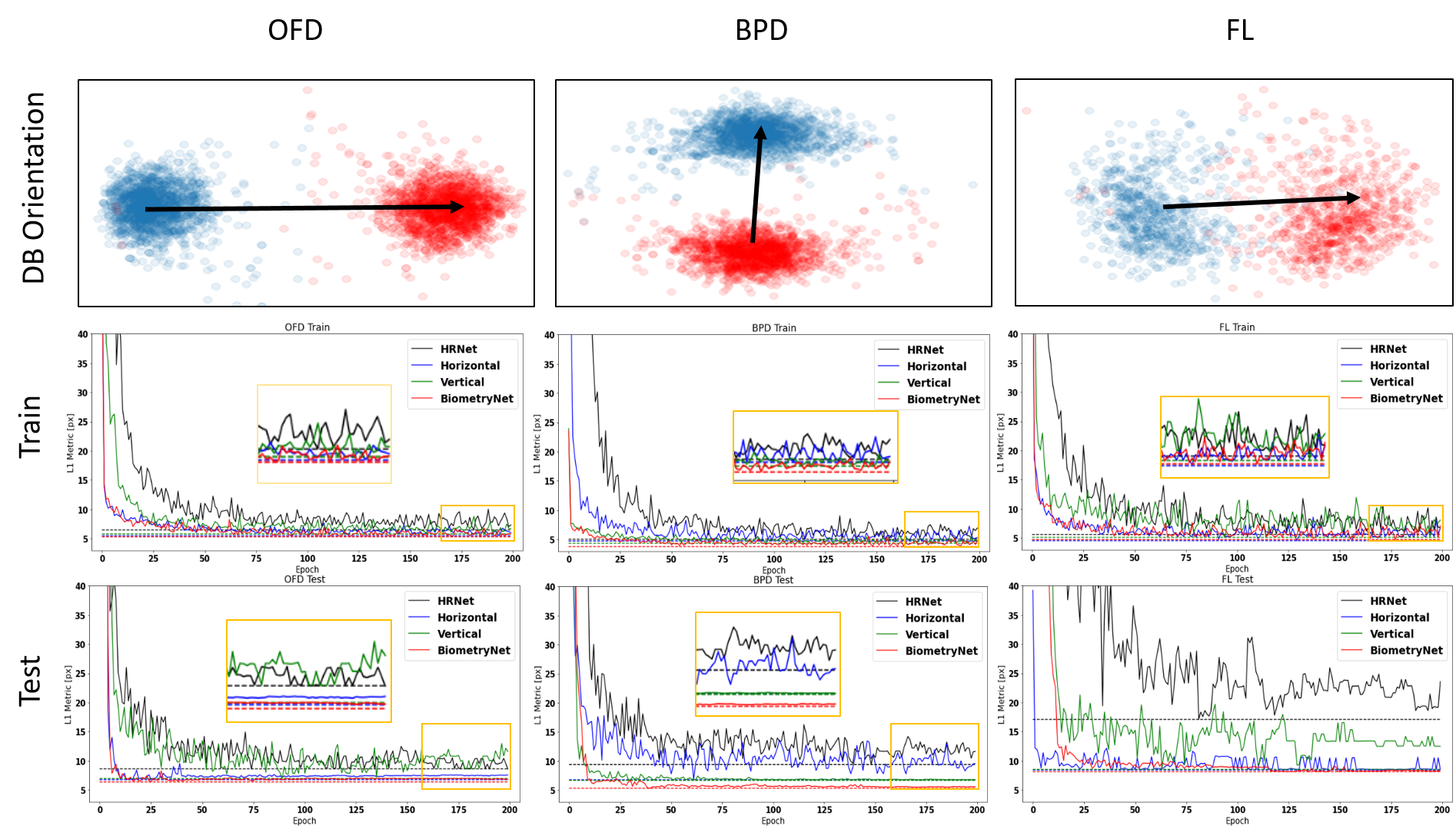}
\caption{Effect of learned dynamic orientation (first row) on training (2nd row) and inference (3rd row) for three measurements (columns) on convergence (yellow boxes): OFD, BPD and FL, in the FP dataset. Four models were trained: \textit{HRNet} (black), with \textit{horizontal} (blue) fixed orientation, \textit{vertical} (green) fixed orientation and \textit{BiometryNet} with DOD (red). Dotted lines denote the best performing epoch metric for each model.  %Learned measurement orientation (first row), L1 metric on model training (second row) and testing (third row)
} \label{fig_orient_exper}
\end{figure}

We conducted three experimental studies on the FP and HC18 datasets to quantify the performance of BiometryNet and compare it to other existing methods. 
%In particular, we examine robustness to variability and generalization capabilities. Performance with respect to the existing methods and validate its capabilities specific to robustness to variabilities and generalization. 
\\[2ex]
{\bf Study 1: Effect of DOD on training:}
we evaluated the effect of DOD on training and inference dynamics on three biometric measurements: OFD, BPD, and FL in the FP dataset. OFD and FL have mostly horizontal orientations and BPD has mostly vertical orientation. Fig.~\ref{fig_orient_exper} shows the results. Note that for all biometric measurements, the learned orientation is similar to the preferred orientation. BiometryNet converged faster and performed better than networks trained on fixed orientations of both the training and test sets, similar to those with the preferred fixed orientation, and better than the Vanilla HRNet, which performed poorly. This advantage becomes more evident when the landmarks' locations are more disperse, e.g. for FL. This shows that DOD provides robustness with respect to orientation variability and yields an improved performance.
\\[2ex]
%we can notice that for all biometries, the learned orientations are the one we expect, in addition, in both train and test sets, DOD achieved the best results, similar to the preferred orientation. Furthermore, we can observe that in all biometries, vanilla HRNet performed poorly, while dynamic tends to converge faster. Thus, we can conclude that using DOD helps the landmark regression network to learn the biometry task.
{\bf Study 2: Performance comparison:} we evaluate the performance of BiometryNet on the FP and HC18 datasets. Table~\ref{tab:expr-table} lists the results. HRNet performed poorly on all biometric measurements, with a high $CI_{95}$ and error. Fixed orientation determination yielded better results in preferred orientation of measurement, i.e. horizontal in OFD and FL, and vertical in BPD. 
FMLNet~\cite{avisdris2021fmlnet} performed better than fixed horizontal in all cases,
% and in FL even was the best performer,
but at the expense of discarding $8\%$ of test inputs in average.
%, 75(8$\%$), 65(9$\%$) and 22(7.5$\%$) images of OFD, BPD and FL, respectively from FP test set, and 55(21$\%$) and 15(5.7$\%$) of OFD and BPD, respectively from HC18 dataset.
BiometryNet outperformed other methods in OFD and BPD biometric measurements in all metrics. While FL has a slightly better bias and ${CI}_{95}$ in FMLNet, considering only included cases by all methods, the paired t-test ($p=0.02$) shows that BiometryNet outperforms all other methods, including FMLNet.

BiometryNet yields a median error (< 0.84mm (OFD), < 0.61mm (BPD) and < 0.62mm (FL)) that is better than the error reported in~\cite{bano2021autofb} (1.30mm (OFD), 0.80mm (BPD) and 2.1mm (FL)). In addition, the results are better than \cite{al2019bpdauto}, which was the best BPD performer in a 2022 review on fetal US biometry \cite{torres2022headus_review}, that achieved a mean error of 2.33mm, bias of 1.49mm and $CI_{95}$ of 5.55mm. Furthermore, the variance of all methods except BiometryNet is higher than the reported inter-observer variability \cite{zhang2017automatic_fhc} of 5.0mm (OFD), 3.0mm (BPD) and 4.3mm (FL) \cite{sarris2012intra}.
%We can note that OFD tends to get higher bias and variance as its landmarks positions tend to be more variable, while mostly being in the shadowed area of the skull. 
These results suggest that overall BiometryNet performs better and is more stable (lower variance) than the existing methods. \looseness=-1 
\begin{table}[tp]
\caption{Study 2 and 3 results for head (OFD, BPD) and femur (FL) biometric measurements for the FP and HC18 datasets. For each biometric measurement, bias, agreement $CI_{95}$, and mean $\overline{L_1}$ and median $\widetilde{L_1}$ differences with respect to expert annotations are listed. The best performance for each metric and dataset in indicated in Bold. (*) indicates only images with reliable predictions were included in FMLNet evaluation.\looseness=-1}
\resizebox{\textwidth}{!}{%
\begin{tabular}{|l|l|l|llll|llll|llll|}
\hline
\textbf{\begin{tabular}[c]{@{}l@{}}Train\\ DB\end{tabular}} &
  \textbf{\begin{tabular}[c]{@{}l@{}}Test\\ DB\end{tabular}} &
  \textbf{Method} &
  \multicolumn{4}{c|}{\textbf{Head - OFD}} &
  \multicolumn{4}{c|}{\textbf{Head - BPD}} &
  \multicolumn{4}{c|}{\textbf{Femur - FL}} \\ \cline{4-15}
\multicolumn{1}{|c|}{} &
  \multicolumn{1}{c|}{} &
  \multicolumn{1}{c|}{} &
  \multicolumn{1}{c|}{\textit{\begin{tabular}[c]{@{}c@{}}Bias\\ {[}mm{]}\end{tabular}}} &
  \multicolumn{1}{c|}{\textit{\begin{tabular}[c]{@{}c@{}}$CI_{95}$\\ {[}mm{]}\end{tabular}}} &
  \multicolumn{1}{c|}{\textit{\begin{tabular}[c]{@{}c@{}}$\overline{L_1}$\\ {[}mm{]}\end{tabular}}} &
  \multicolumn{1}{c|}{\textit{\begin{tabular}[c]{@{}c@{}}$\widetilde{L_1}$\\ {[}mm{]}\end{tabular}}} &
  \multicolumn{1}{c|}{\textit{\begin{tabular}[c]{@{}c@{}}Bias\\ {[}mm{]}\end{tabular}}} &
  \multicolumn{1}{c|}{\textit{\begin{tabular}[c]{@{}c@{}}$CI_{95}$\\ {[}mm{]}\end{tabular}}} &
  \multicolumn{1}{c|}{\textit{\begin{tabular}[c]{@{}c@{}}$\overline{L_1}$\\ {[}mm{]}\end{tabular}}} &
  \multicolumn{1}{c|}{\textit{\begin{tabular}[c]{@{}c@{}}$\widetilde{L_1}$\\ {[}mm{]}\end{tabular}}} &
  \multicolumn{1}{c|}{\textit{\begin{tabular}[c]{@{}c@{}}Bias\\ {[}mm{]}\end{tabular}}} &
  \multicolumn{1}{c|}{\textit{\begin{tabular}[c]{@{}c@{}}$CI_{95}$\\ {[}mm{]}\end{tabular}}} &
  \multicolumn{1}{c|}{\textit{\begin{tabular}[c]{@{}c@{}}$\overline{L_1}$\\ {[}mm{]}\end{tabular}}} &
  \multicolumn{1}{c|}{\textit{\begin{tabular}[c]{@{}c@{}}$\widetilde{L_1}$\\ {[}mm{]}\end{tabular}}} \\ \hline
\multirow{6}{*}{\textbf{FP}} &
  \multirow{5}{*}{\textbf{FP}} &
  \textit{HRNet~\cite{WangSCJDZLMTWLX19}} &
  \multicolumn{1}{r|}{6.23} &
  \multicolumn{1}{r|}{26.40} &
  \multicolumn{1}{r|}{6.30} &
  \multicolumn{1}{r|}{3.30} &
  \multicolumn{1}{r|}{2.84} &
  \multicolumn{1}{r|}{22.57} &
  \multicolumn{1}{r|}{3.20} &
  \multicolumn{1}{r|}{0.80} &
  \multicolumn{1}{r|}{1.80} &
  \multicolumn{1}{r|}{18.40} &
  \multicolumn{1}{r|}{2.70} &
  \multicolumn{1}{r|}{0.62} \\ \cline{3-15} 
 &
   &
  \textit{Horizontal} &
  \multicolumn{1}{r|}{2.65} &
  \multicolumn{1}{r|}{10.23} &
  \multicolumn{1}{r|}{2.87} &
  \multicolumn{1}{r|}{1.90} &
  \multicolumn{1}{r|}{2.36} &
  \multicolumn{1}{r|}{21.60} &
  \multicolumn{1}{r|}{2.78} &
  \multicolumn{1}{r|}{0.76} &
  \multicolumn{1}{r|}{0.17} &
  \multicolumn{1}{r|}{3.27} &
  \multicolumn{1}{r|}{0.99} &
  \multicolumn{1}{r|}{\textbf{0.59}} \\ \cline{3-15} 
 &
   &
  \textit{Vertical} &
  \multicolumn{1}{r|}{4.73} &
  \multicolumn{1}{r|}{23.51} &
  \multicolumn{1}{r|}{4.86} &
  \multicolumn{1}{r|}{2.46} &
  \multicolumn{1}{r|}{0.77} &
  \multicolumn{1}{r|}{8.28} &
  \multicolumn{1}{r|}{1.28} &
  \multicolumn{1}{r|}{0.65} &
  \multicolumn{1}{r|}{0.33} &
  \multicolumn{1}{r|}{10.27} &
  \multicolumn{1}{r|}{1.47} &
  \multicolumn{1}{r|}{0.65} \\ \cline{3-15} 
 &
   &
  \textit{FMLNet*~\cite{avisdris2021fmlnet}} &
  \multicolumn{1}{r|}{1.96} &
  \multicolumn{1}{r|}{7.80} &
  \multicolumn{1}{r|}{2.16} &
  \multicolumn{1}{r|}{1.43} &
  \multicolumn{1}{r|}{1.30} &
  \multicolumn{1}{r|}{14.56} &
  \multicolumn{1}{r|}{1.71} &
  \multicolumn{1}{r|}{0.65} &
  \multicolumn{1}{r|}{\textbf{0.14}} &
  \multicolumn{1}{r|}{\textbf{3.00}} &
  \multicolumn{1}{r|}{1.02} &
  \multicolumn{1}{r|}{0.68} \\ \cline{3-15} 
 &
   &
  \textit{\textbf{BiometryNet}} &
  \multicolumn{1}{r|}{\textbf{0.21}} &
  \multicolumn{1}{r|}{\textbf{2.75}} &
  \multicolumn{1}{r|}{\textbf{1.01}} &
  \multicolumn{1}{r|}{\textbf{0.71}} &
  \multicolumn{1}{r|}{\textbf{0.04}} &
  \multicolumn{1}{r|}{\textbf{2.50}} &
  \multicolumn{1}{r|}{\textbf{0.77}} &
  \multicolumn{1}{r|}{\textbf{0.58}} &
  \multicolumn{1}{r|}{0.18} &
  \multicolumn{1}{r|}{3.03} &
  \multicolumn{1}{r|}{\textbf{0.97}} &
  \multicolumn{1}{r|}{0.62} \\ \cline{2-15} 
 &
  \textbf{HC18} &
  \textit{\textbf{BiometryNet}} &
  \multicolumn{1}{r|}{2.31} &
  \multicolumn{1}{r|}{5.21} &
  \multicolumn{1}{r|}{2.46} &
  \multicolumn{1}{r|}{1.85} &
  \multicolumn{1}{r|}{0.84} &
  \multicolumn{1}{r|}{2.70} &
  \multicolumn{1}{r|}{1.06} &
  \multicolumn{1}{r|}{0.91} &
   &
   &
   &
   \\ \cline{2-11}
\multirow{6}{*}{\textbf{HC18}} &
  \multirow{5}{*}{\textbf{HC18}} &
  \textit{HRNet~\cite{WangSCJDZLMTWLX19}} &
  \multicolumn{1}{r|}{0.64} &
  \multicolumn{1}{r|}{6.01} &
  \multicolumn{1}{r|}{1.51} &
  \multicolumn{1}{r|}{0.92} &
  \multicolumn{1}{r|}{2.64} &
  \multicolumn{1}{r|}{21.48} &
  \multicolumn{1}{r|}{3.10} &
  \multicolumn{1}{r|}{0.71} &
   &
   &
   &
   \\ \cline{3-11}
 &
   &
  \textit{Horizontal} &
  \multicolumn{1}{r|}{2.82} &
  \multicolumn{1}{r|}{23.9} &
  \multicolumn{1}{r|}{3.69} &
  \multicolumn{1}{r|}{0.93} &
  \multicolumn{1}{r|}{1.35} &
  \multicolumn{1}{r|}{12.86} &
  \multicolumn{1}{r|}{1.75} &
  \multicolumn{1}{r|}{\textbf{0.59}} &
   &
   &
   &
   \\ \cline{3-11}
 &
   &
  \textit{Vertical} &
  \multicolumn{1}{r|}{4.02} &
  \multicolumn{1}{r|}{29.15} &
  \multicolumn{1}{r|}{4.92} &
  \multicolumn{1}{r|}{0.97} &
  \multicolumn{1}{r|}{0.50} &
  \multicolumn{1}{r|}{5.13} &
  \multicolumn{1}{r|}{0.98} &
  \multicolumn{1}{r|}{0.65} &
   &
   &
   &
   \\ \cline{3-11}
 &
   &
  \textit{FMLNet*~\cite{avisdris2021fmlnet}} &
  \multicolumn{1}{r|}{2.23} &
  \multicolumn{1}{r|}{17.48} &
  \multicolumn{1}{r|}{2.61} &
  \multicolumn{1}{r|}{1.02} &
  \multicolumn{1}{r|}{0.73} &
  \multicolumn{1}{r|}{4.00} &
  \multicolumn{1}{r|}{0.93} &
  \multicolumn{1}{r|}{0.64} &
   &
   &
   &
   \\ \cline{3-11}
 &
   &
  \textit{\textbf{BiometryNet}} &
  \multicolumn{1}{r|}{\textbf{0.56}} &
  \multicolumn{1}{r|}{\textbf{4.43}} &
  \multicolumn{1}{r|}{\textbf{1.39}} &
  \multicolumn{1}{r|}{\textbf{0.84}} &
  \multicolumn{1}{r|}{\textbf{0.16}} &
  \multicolumn{1}{r|}{\textbf{3.54}} &
  \multicolumn{1}{r|}{\textbf{0.88}} &
  \multicolumn{1}{r|}{0.61} &
   &
   &
   &
   \\ \cline{2-11} 
 &
  \textbf{FP} &
  \textit{\textbf{BiometryNet}} &
  \multicolumn{1}{r|}{-3.24} &
  \multicolumn{1}{r|}{6.01} &
  \multicolumn{1}{r|}{3.54} &
  \multicolumn{1}{r|}{2.72} &
  \multicolumn{1}{r|}{-1.11} &
  \multicolumn{1}{r|}{3.35} &
  \multicolumn{1}{r|}{1.40} &
  \multicolumn{1}{r|}{1.08} &
   &
   &
   &
   \\ \hline
\end{tabular}%

}
\label{tab:expr-table}
\end{table}
\\[2ex]
{\bf Study 3: Impact of training dataset:} to demonstrate the generalization capabilities of BiometryNet, we analyzed its performance by cross-validation on unseen datasets. For this purpose, we train the OFD and BPD models on FP dataset and test each on HC18 dataset, and vice versa. Table~\ref{tab:expr-table} lists the results. Note that training on one dataset and testing on the other (rows 6 and 12) results in high mean and median errors, with significant and complementary (2.31mm vs -3.24mm in OFD, 0.81mm vs -1.11mm in BPD) bias and acceptable variance. In contrast, testing on the same dataset (rows 5 and 11) results in low bias, but similar variance ($CI_{95}$) as before. This bias can be explained by the differences in annotation protocol between the two datasets. In HC18, the landmark annotations lie between the outer contours of the skull, while in FP, they are marked in the middle of the fetal skull contour, thus resulting in a consistent bias (Fig. 2 supp.). Clinically, both protocols are acceptable, and their selection depends on the specific clinical site. 
We also observe that using FP annotations for network training produced better and more consistent results than using those of HC18. This may occur because HC18 annotations are extracted from HC ellipse rather than annotated directly at the BPD/OFD landmarks.
We conclude from these results that BiometryNet is capable of generalization, as it can learn to annotate like the annotation protocol, and can be tuned for use across many sites and protocols. 
% BiometryNet demonstrated generalizability across different datasets, as it learned to annotate like the annotation protocol. Thus, BiometryNet can be tuned for use across many different clinical sites and protocols.
%While \cite{avisdris2021fmlnet} performed plane selection and biometry in fetal MRI in two seperated networks, \cite{bano2021autofb} tried to automate both tasks in the same network by using multi-class segmentation, but showed lower performance in some cases. In US fetal biometry,  \cite{burgos2020zdataset} showed that fetal plane selection can be automated using CNN in high accuracy and may be used in an integrated framework. In addition, evaluation of abdominal plane and its measurements ... , similar to \cite{bano2021autofb} 
%\\[2ex]
%{\bf Future work:} 
% \\[2ex]
% As for {\bf future work}, we will analyze the abdominal plane and its measurement using BiometryNet. Furthermore, BiometryNet assumes SP is already available for the measurement, which is not always the case in clinical settings. Therefore, we will design a holistic network to jointly learn SP selection and fetal biometry. % from all three SPs.
\section{Conclusions}
We proposed BiometryNet, an end-to-end network for automatic fetal biometry estimation from standard US planes. BiometryNet only required anatomical landmarks annotations for biometric measurements prediction without the need of any geometric methods as post-processing. To overcome the variability that is inherently present in the acquired US planes, we introduced a novel dynamic orientation determination mechanism which enforced measurement-wise orientation consistency in network training. This resulted in reduced variability and improved landmarks' localization, thus leading to more accurate biometric measurements compared to the state-of-the-art methods. Through the analysis of two large and independent datasets, we demonstrated the generalization ability of our proposed method. Moreover, BiometryNet resulted in minimal error which is lower than the clinically permissible error \cite{sarris2012intra}, thus showing the potential for clinical translation to improve fetal growth assessment.
\\[1ex]
In future work, we plan to analyze the abdominal plane and its measurement using BiometryNet. Furthermore, BiometryNet assumes SP is already available for the measurement, which is not always the case in clinical settings. Therefore, we will design a holistic network to jointly learn SP selection and fetal biometry. % from all three SPs. 
\\[1ex]
\textbf{Acknowledgements} This research was partly supported by the Wellcome/EPSRC Centre for Interventional and Surgical Sciences (WEISS) [203145/Z/16/Z]; the Engineering and Physical Sciences Research Council (EPSRC) [EP/P027938/1, EP/R004080/1, EP/P012841/1]; the Royal Academy of Engineering Chair in Emerging Technologies Scheme, and Horizon 2020 FET Open [863146]; Kamin Grants [63418, 72126] from the Israel Innovation Authority.

\bibliographystyle{splncs04}
\bibliography{main.bib}

\newpage
% This is samplepaper.tex, a sample chapter demonstrating the
% LLNCS macro package for Springer Computer Science proceedings;
% Version 2.20 of 2018/03/10
%

%
%\title{Landmark-based Automatic Ultrasound Fetal Biometry}
\title{BiometryNet: Landmark-based Fetal Biometry Estimation from Standard Ultrasound Planes (Supplementary Material)}

\author{Netanell Avisdris \textsuperscript{\Letter} \inst{1,2} \and
Leo Joskowicz\inst{1} \and
Brian Dromey\inst{4,6} \and
Anna L. David\inst{6} \and
Donald M. Peebles\inst{6} \and
Danail Stoyanov\inst{4,5} \and
Dafna Ben Bashat\inst{2,3} \and
Sophia Bano\inst{4,5} }

\authorrunning{N. Avisdris et al.}

\institute{School of Computer Science and Engineering, The Hebrew U. of Jerusalem, Israel \email{\{netana03,josko\}@cs.huji.ac.il} \and
Sagol Brain Institute, Tel Aviv Sourasky Medical Center, Israel \and Sagol School of Neuroscience and Sackler Faculty of Medicine, Tel Aviv U., Israel \and
Wellcome/EPSRC Centre for Interventional and Surgical Sciences(WEISS),
University College London, London, UK \and
Department of Computer Science, University College London, London, UK \and
Elizabeth Garrett Anderson Institute for Women’s Health, University College London, London, UK
}

\maketitle              % typeset the header of the contribution

\begin{figure}[ht]
\includegraphics[width=\textwidth]{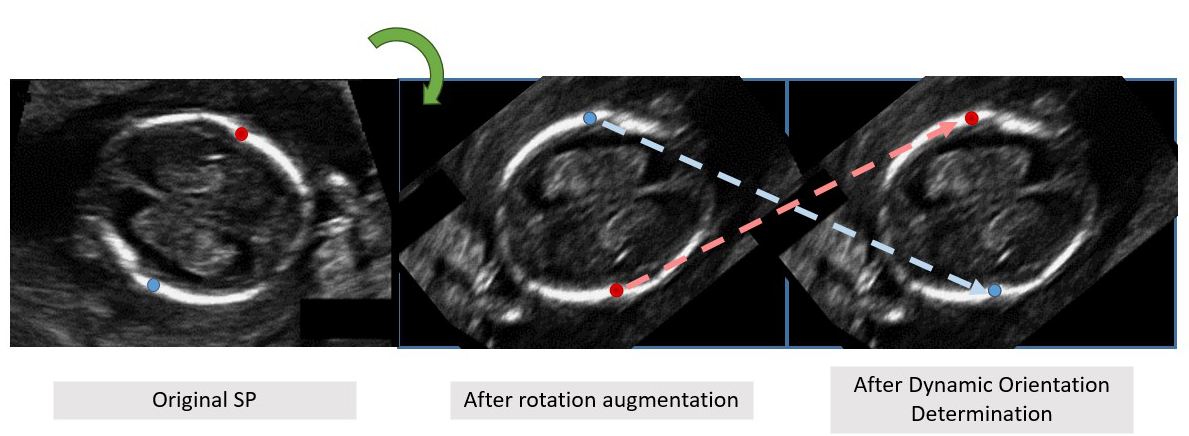}
\caption{Dynamic Orientation Determination (DOD): illustrative example. (a) original standard plane (SP) ultrasound (US) image of fetal trans-ventricular plane with the bi-parietal diameter (BPD) biometric measurement up (red) and down (green) ground truth landmarks; (b) SP after rotation showing an inconsistent landmark labeling in which the up and down landmarks are switched); (c) reassignment of the landmark up and down labels by DOD (blue and red intermittent lines). } \label{fig_dodmotivation}
\end{figure}

\begin{figure}[htp]
\includegraphics[width=\textwidth]{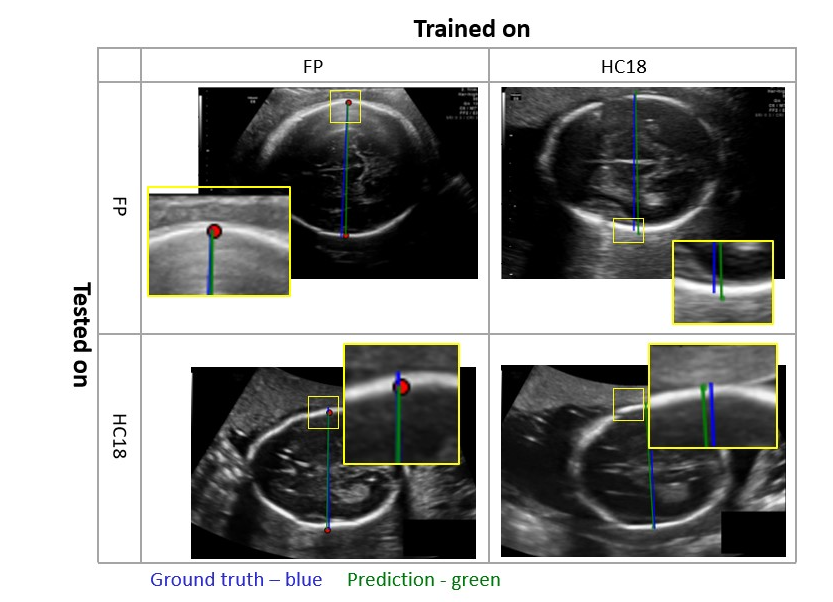}
\caption{Example showing different results of two annotation methods and the resulting BiometryNet predictions. 
Two models were trained, one on the FP training dataset (first column) and the second on the HC18 training dataset  (second column). Both were run on the FP testing dataset (first row) and the HC18 testing dataset (second row). Landmarks are marked in the middle of the fetal skull in FP dataset and on the outer extrema of the skull in HC18 dataset. This introduces a bias that results in a relatively larger error, although still within the clinically acceptable limit when generalizing from FP$\rightarrow$HC18 or HC18$\rightarrow$FP. The quantitative results are presented in Table 1 and discussed in Study 2 and 3. The images show ground truth annotations (blue) and computed predictions (green) on representative examples of BPD biometry. The relevant areas are enlarged in the yellow box. } \label{fig_pipeline2}
\end{figure}

\end{document}